# Title of Manuscript

Direction of slip modulates the perception of slip distance and slip speed

# List of Authors

1. Ayesha Tooba Khan[1] (ORCID: 0000-0002-0690-8517)
2. Deepak Joshi[1,2] (ORCID: 0000-0002-2438-3232)
3. Biswarup Mukherjee[1,2] (ORCID: 0000-0001-5528-3763)

   **Corresponding Author:** Biswarup Mukherjee, Email: bmukherjee@cbme.iitd.ac.in

**Affiliation**

1: Centre for Biomedical Engineering, Indian Institute of Technology Delhi, New Delhi, India.

2: Department of Biomedical Engineering, All India Institute of Medical Sciences, New Delhi, India.

**Acknowledgements:**

The authors warmly thank all the participants who participated in the study and willingly gave us several hours of their time. We also thank the Central Research Facility, Indian Institute of Technology Delhi, New Delhi for providing 3D printing and Laser cutting services to develop the device.

**Funding:** This work was supported in part by the Science & Engineering Research Board, Department of Science and Technology, Government of India through a Core Research Grant (CRG/2021/004967, PI: B. Mukherjee) and by IIT Delhi through a New Faculty Seed Grant (PLN12/03BM, PI: B. Mukherjee).

# Direction of slip modulates the perception of slip distance and slip speed

## Abstract

**Purpose:** The purpose of this study was to investigate the psychophysical understanding of the slip stimulus. We emphasized that the perception of slip and its characteristics, such as slip distance and slip speed depend on the interaction between slip direction, slip distance as well as slip speed.

**Methods:** We developed a novel slip induction device to simulate the artificial sense of slip. We conducted a psychophysical experiment on eight healthy subjects. The experiment was designed to evaluate the effect of slip direction on slip perception as well as on the perception of slip distance and slip speed. A series of psychophysical questions were asked at the end of the slip stimulation to record the subjective responses of the participants. The average success rate (%) was used to quantify the subject responses.

**Results:** We demonstrated that the perception of slip is independent of slip direction however, perception of slip distance and slip speed are significantly modulated by slip direction. We also observed that a significant interaction exists between slip distance and slip speed in the upward slip direction. It was also observed that the average success rate was significantly different for various combinations of slip distance and slip speed in the upward slip direction.

**Conclusions:** Our study clearly establishes a significant interaction between the slip direction, slip distance, and slip speed for psychophysical understanding of the perception of slip distance and slip speed.

## I. Introduction

The sense of touch and the ability to perceive object slippage involve complex interactions between sensory receptors, neural pathways, and the central nervous system (CNS), forming a closed-loop control system. It relies on both the forward kinematics of the musculoskeletal system as well as the sensing mechanisms of the somatosensory system. Thus, able-bodied individuals modulate grip force during object manipulation according to the somatosensory feedback from the fingertips and the proprioceptive feedback from muscle spindles adapting based on the frictional condition between the fingertips and the object [1]. However, during object slippage, an involuntary response initiated by the spinal cord comes into action. This response is faster than voluntary corrective action as it bypasses the brain while initiating the corrective motor command. On the other hand, individuals with neuromuscular deficits, such as those with upper limb amputation, paralysis following spinal cord injury, or stroke, may not be able to initiate fast, involuntary reactions due to disruptions in efferent and afferent CNS volleys, leading to object slippage due to improper modulation of the grip force [2]. This establishes the importance of slip detection as well as slip perception while considering the slip characteristics such as slip direction, slip speed, and slip distance to apply the appropriate corrective measures in the artificial closed loop system of individuals with neuromuscular deficits to prevent object slippage.

Detection and perception of slip has been an integral research arena explored in various neurophysiological, biomechanical, and psychophysical studies. Previous research has shown that Pacinian corpuscles are mainly responsible for encoding the intensity of frictional forces for the detection of slip. In contrast, Meissner's corpuscles, Merkel's cells, and Ruffini endings play a crucial role in determining the slip direction. A recent study demonstrated a fast, automatic coupling arising between the slip sensation and the arm response when the arm was moved in the slip direction [3]. In the same study, fast feedback responses arising from slip stimuli were significantly different for slip stimuli with different slip speeds having the same slip distance but not for slip stimuli with different slip distances having the same slip speed. In tandem with the neurophysiological basis of slip detection, researchers have also investigated the biomechanical deformation at the fingertips during tangential loading in different directions, which exhibited complex mechanics. An increased tangential force led to a reduced contact area between the surface and the fingertip, thereby suggesting the nonlinear stiffening of the skin [4]. The temporal evolution of surface strain revealed strain wave patterns propagating from the peripherals towards the centre of the contact area [5]. Various slip stages have also been explored quantitatively using stick ratio and how the central nervous system perceives them to modulate the grip force [6]. Delhaye et al. also studied the activity of single tactile afferents as a response to slip using microneurography and mapped their response with surface deformations at the fingertips[7]. They found dominant contributions of Meissners' corpuscles in perceiving compressive strain rates and, hence, detecting the partial slips that can modulate grip force during object manipulation [7]. André et al. investigated the effect of skin hydration on fingertip dynamics during various slip stages. They found a linear relationship between the stick ratio and the load force, while the grip force was found to be more unstable under dry skin conditions [8]. Additionally, a systematic psychophysical investigation of how individuals perceive and respond to slip events is crucial for exploring the cognitive and subjective dimensions involved in slip perception. Although a very limited number of psychophysical studies have been done to understand slip perception, these studies have revealed that the perception of slip depends upon slip characteristics such as surface texture [9], slip speed, slip distance, and slip direction [6, 9]. For example, horizontal slip direction (lateral-medial or medial-lateral) viz., slip in the transverse plane has been found to be significantly impacting the subjects' responses to slip perception by Srinivasan et al. [9] whereas, a recent study by Barrea et al. shows no significant effect of horizontal slip stimulation was observed by Barrea et al. [6].

Based on this study, we envisaged that it is not only the slip direction but rather the interaction between the slip direction, slip distance and slip speed which determines the perception of the slip at the higher-order cognitive levels. It further raises a pertinent question of whether a slip distance and a slip speed are perceived identically in both the slip directions. Moreover, no studies were found to investigate either the neurophysiological or the psychophysical behaviour of the slip perception in the vertical slip direction, which closely mimics the real-life scenario of object slippage from our hands. Additionally, all the studies done in the past to either study slip perception [6] or slip detection [4, 7, 9] have used either the pinch grasp or the single index finger. However, in everyday life, we usually use prehension grip to perform activities of daily living. Therefore, it is crucial to investigate the psychophysical, neuromuscular, or biomechanical manifestation of object slippage using a prehension grasp. In this work, we developed a slip induction device capable of simulating the artificial sense of object slippage in the vertical direction (upwards and downwards) while the participant holds on to the grasping plates of the device using a prehension grip. We conducted a psychophysical experiment with able-bodied individuals to develop a cognitive understanding of the interaction between the slip direction, slip distance and slip speed. Participants were provided with either an upward or a downward slip stimulus. We have considered five different slip distances and slip speeds in upward as well as downward directions, thereby giving fifty unique combinations of slip stimuli. Thus, our research provides the first investigation of the psychophysical behaviour of interaction between slip speeds, slip distances and slip direction during object manipulation.

## II. Materials and Methods

Our study aims to investigate the influence of slip speed, distance, and direction on the perception of object slip. We hypothesized that:

A: There is no significant effect of slip direction on slip perception.

B: There is a significant effect of slip direction on the perception of magnitudes of slip distance and slip speed.

C: The interaction between slip distance and slip speed significantly affects the perception of magnitudes of slip distance and slip speed.

The following sections describe the experimental setup and analysis to test the hypotheses stated above.

### A. Participants

Eight healthy human participants (aged 27.25±4.83, six males, two females) with no history of neuromuscular disorder, surgical intervention in the upper extremity, or physical or cognitive impairment were recruited to perform the psychophysics experiment to perceive the artificial sense of object slippage. All the participants reported to be right hand dominant and had normal or corrected to normal vision. The Institute Ethics Committee at the Indian Institute of Technology Delhi approved all experimental procedures (Ref no: 2021/P052). All participants provided informed consent to participate in the experiment.

### B. Design and development of slip induction device

We have developed the slip induction device (SID) shown in Figure 1. (b), which was used to simulate the artificial sense of object slippage. The device is a linear actuator driven by a stepper motor (NEMA17, 4.8 kg-cm) controlled by a two-phase hybrid stepper motor driver (TB6600HG, Toshiba Corp., Japan) with a current of 2.0 A/phase and 32 micro-step mode. The device is equipped with the force measurement assembly comprising the four load cells to measure the load and the grip forces exerted by the fingers and the thumb on either side of the grasping plates. Grasping plates constructed from transparent acrylic sheets are attached to the load cells on one side while the subject has to grasp the other side during the experiment. The grip aperture of the device is approximately 60 mm. An optical quadrature rotary encoder (3806-OPTI-600-AB-OC, Orange, China) was coupled to the stepper motor to measure the rotational speed of the SID. A microcontroller (Teensy 3.2, PJRC INC., USA) development board is used to control the overall mechanism of the device and communicate with a PC in real time. Custom-designed parts, including the base holding the stepper motor, the top holding the encoder, the grasping plates carriage, the grasping plate mounts, and the limit switch holder of the SID, were rapid prototyped using fused deposition modelling 3D printer (Motion System and Tool Changer, E3D, UK).

### C. Development of software interface

We developed a MATLAB app-based user interface (MATLAB 2023b, The MathWorks, Inc., USA) to conduct the experiment using the SID. The front end of the interface had the experimenter console and the user console. The demographics of the participants were recorded at the experimenter console. Then a familiarization session comprising 11 command sequences was given to the participant to allow the participant to be acquainted with the interface. Before starting the session, the device was homed to its zero vertical position. Then, baseline data from the four load cells during the idle state (when the participant has not grasped the grasping plates of the device) was recorded for 20 s during the visualisation of grip force. The experimenter manually commenced the trials after the completion of the familiarization session. The interface was programmed to randomize the sequence of the stimuli presented to the user during each block. Each trial had three phases, viz., the calibration phase, the slip simulation phase, and the perceptual judgement phase, as shown in Figure 1. (d). During the calibration phase, the user console showed the instantaneous grip force values exerted by the participant while grasping the device (Refer Figure 1. (c)). This was done to maintain a standardized grip force at the time of slip stimulation to avoid any bias arising due to different grip forces exerted by the participant on the grasping plates. After each trial, psychophysics questions would be posed to the participant.

### D. Experimental Paradigm

To test the hypotheses, we designed the experiment protocol to assess the effect of slip direction on the slip perception as well as the slip characteristics, viz., slip distance and slip speed. We considered five values of slip distance from 2-10 mm in increments of 2 mm and five values for slip speed from 2-10 mm/s in increments of 2 mm/s. This resulted in twenty-five unique combinations of slip distance and slip speed forming the command sequence to be given to the SID, as shown in Figure 1. (d). The SID was programmed to simulate the slip alternatively in the upward direction and then in the downward direction with the same slip characteristics. It is important to note that slip direction was not randomized to enable comparison of the same slip stimulus between upward and downward directions. Then, the psychophysical questions were asked to compare the slip distance and slip speed of the current and previous trials. Additionally, six catch trials, with a duration of 5 seconds each, were added to ensure no learning effects. The catch trials were interleaved such that the first and the last trials were always catch trials. Thus, effectively, there were 56 randomised trials (2 directions x 5 slip distance x 5 slip speed and six catch trials) forming one command sequence for one block. Four uniquely randomised command sequences were generated for four blocks to minimize ordering effects.

The participant was required to be seated comfortably in a chair with their dominant hand resting on the padded armrest and the forearm of the participant as restrained using Velcro bands as shown in Figure 1. (a). The participant was instructed to hold the grasping plates of the SID such that the angle subtended by the long axis of the humerus and the long axis of the radius in the sagittal plane was maintained at 135°. To avoid biased psychophysical responses due to the visual or auditory cues arising from the movement or mechanical noise of the stepper motor, the visual feedback of the device to the participant was occluded with a black curtain between the monitor and the device.

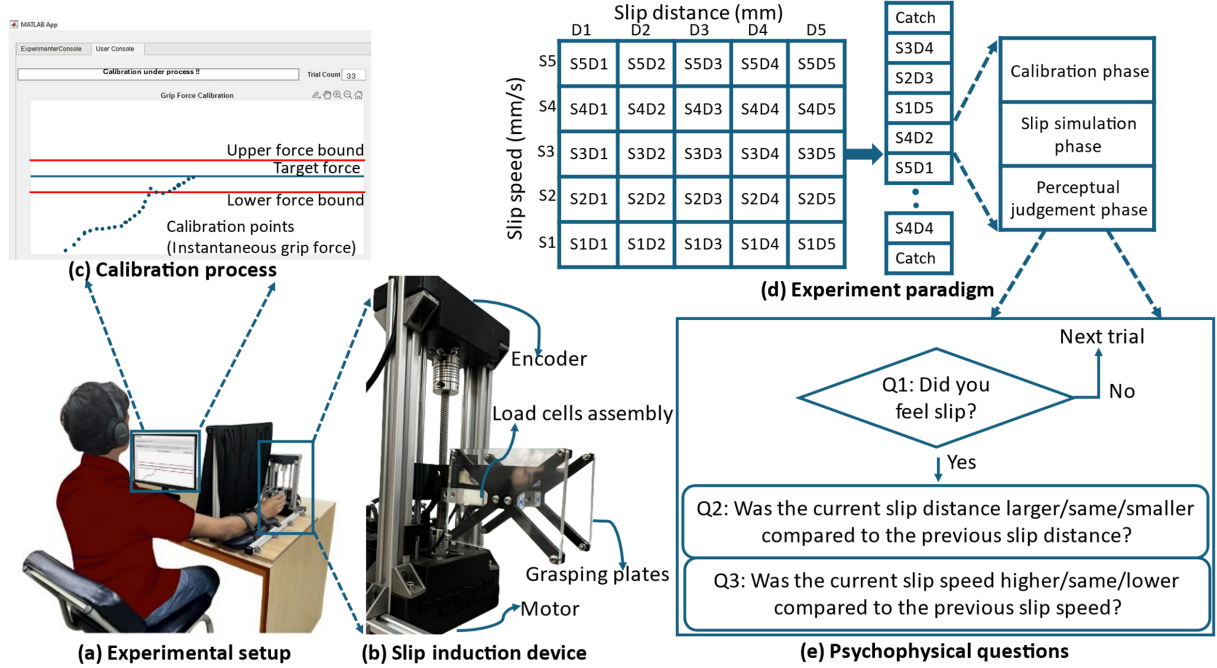

Figure 1. (a) Experimental setup showing the participant performing the task on the slip inducing device (SID). (b) The enlarged view of the SID showing the load cell, linear stage, and grasping apparatus. (c) The enlarged view of the user console during the calibration process showing the target force and the user exerted force reading. (d) Generation of randomized command sequence for various combinations of slip speeds and distances. (e) Psychophysical questions asked from the participant during the perceptual judgement phase.

Auditory feedback of the device to the participant was minimized by playing white noise in the headphones (Bose QuietComfort 25, Bose Corp., USA) with active noise cancellation. The task was performed on a PC (Windows 10, 64-bit, core i7 CPU, 3.2 GHz processor with 16 GB RAM) running the interface described in the previous section. In each trial, there were three phases: the calibration phase, the slip simulation phase, and the perceptual judgement phase. During the calibration phase, the participant was asked to maintain the grip force exerted by the fingers between 1.5±0.1 N for 125 ms. This was done to allow the user to prepare for the oncoming stimulus if and to avoid any bias in slip perception arising because of the difference in the initial grip force of the participant. If the participant succeeded in maintaining this force within 6 s, the trial proceeded to the slip simulation phase; otherwise, the trial was marked as a failed trial. The visual feedback of the instantaneous grip force along with the target force bounds was provided to the participant during the calibration phase. Following successful calibration, the participants were presented with a slip stimulus by the SID. At the end of each successful trial during the perceptual judgement phase, the participant was asked the following questions:

Q1: Did you feel the slip?

Q2: Was the current slip distance larger/same/smaller compared to the previous slip distance?

Q3: Was the current slip speed higher/same/lower compared to the previous slip speed?

Q2 and Q3 were asked only if an affirmative response was recorded for Q1. The entire flow of the questions conditional to specific responses is shown in the flow chart shown in Figure 1. (e). After each block, the participant was provided a rest of 10 min. The total duration of the study was approximately 3 hrs.

## E. Performance evaluation:

In each block, a total of 56 unique stimuli were presented to each participant. Thus, for each unique stimulus, there were four responses recorded during the four blocks. We analyzed the responses of the participants to the perceptual judgment questions and compared them to the presented stimulus. The percentage of trials for which the participant responded correctly for each unique stimulus was considered as the success rate of the participant corresponding to that stimulus.

$$Success\ rate = \frac{Correct\ responses\ in\ our\ sessions}{Total\ number\ of\ sessions} \times 100\ \%$$

The success rate for eight participants was averaged across the participants, and color-coded plots were plotted for visual understanding of the outcomes. The color intensity reflects the average success rate mapped corresponding to a particular slip distance (mm) and slip speed (mm/s). The standard deviation resulting from average participants' responses is mentioned over the corresponding cell of the color-coded matrix plot. The analysis was performed in a custom-developed script in MATLAB.

## F. Statistical analysis

We performed a three-way repeated measures analysis of variance (ANOVA) to determine if a statistically significant three-way interaction exists between the slip direction, slip distance, and slip speed on the perception of slip, perception of slip distance, and perception of slip speed. The test was performed with the alternative that there exists the effect of slip direction on the perception of slip distance and the slip speed but not on the perception of slip. We further performed the follow-up tests (simple two-way interactions and simple-simple main effects) if a significant three-way interaction was found between the independent variables, viz., slip direction, slip distance, and slip speed. We considered the Greenhouse-Geisser correction if Mauchley's Test of sphericity showed a violation of the assumption of sphericity in any case. All statistical analyses were performed using IBM SPSS Statistics (Version 28.0, IBM Corp, Armonk, USA).

# III. Results

The participant was provided with a slip stimulus having specific slip characteristics (slip distance and slip speed) first in the upward direction, then in the downward direction with the same slip characteristics for two consecutive trials. If the participant succeeded in maintaining the grip force within the prescribed range of 1.5±0.2 N for a duration of 125 ms within 6 s, slip simulation began, followed by the psychophysical questions posed to the participant, thereby making it a successful trial. The average percentage of successful trials for eight participants was 88.62±12.1 %. After each trial, the participant had to respond if they felt a slip during the slip stimulation phase. The average success rate corresponding to various slip distances and slip speeds for the participants' responses to this question is shown in Figure 2. (a and b) for upward and downward slips, respectively. If the participant's response to the above question was 'yes', he/she was further asked to differentiate the magnitudes of slip distance and slip speed of the current trial from the previous trial. The average success rate corresponding to various slip distances and slip speeds for the perception of difference in slip distances is shown in Figure 2. (c and d), while Figure 2. (e and f) show the average success rate for the perception of difference in slip speeds. The average success rate mentioned corresponding to 0 mm slip distance and 0 mm/s slip speed is the average participants' response for the six catch trials. For ease of representation, we will report the intext results in the following format: average success rate ± standard deviation (slip distance (mm), slip speed (mm/s)). For example, for catch trials, the average success rate will be reported as 95.8±8.1% (0 mm, 0 mm/s).

**I) Effect of slip direction on the perception of slip**

To validate our first hypothesis, we analyzed the participants' responses to the first psychophysics question, i.e., Q1. Slip stimuli in the upward direction resulted in higher average success rates except for slip stimuli with smallest slip speed of 2mm/s (90.6±18.6% (4mm,2mm/s), 93.8±11.6% (2mm,2mm/s)) and with largest slip distance (10mm) having moderate slip speed of 6mm/s (90.6±12.9% (10mm,6mm/s)). For the downward slip direction, Figure 2. (b) shows that the perception slip was diminished for smaller slip distances and lower slip speeds (2mm/s and 4mm/s) (90.6±18.6% (2mm,2mm/s), 87.5±18.9% (4mm,2mm/s), and 90.6±18.6% (4mm,4mm/s)) as compared to slip stimuli with higher distances and speeds except for one slip stimulus where the average success rate was found to be lowest (81.2±17.7% (10mm,6mm/s)).

Moreover, as evident from Figure 3., the overall average success rate, i.e., average success rate for all slip distances and slip speeds taken together, for perception of slip in the upward direction was found to be 98.38±2.87 %, while for the downward direction, it was found to be 95.38±4.34 %. The three-way repeated measures ANOVA reveals that the three-way interaction between the slip direction, slip distance and slip speed was not statistically significant for the perception of slip (F (16,112) = 0.597, p = 0.881).

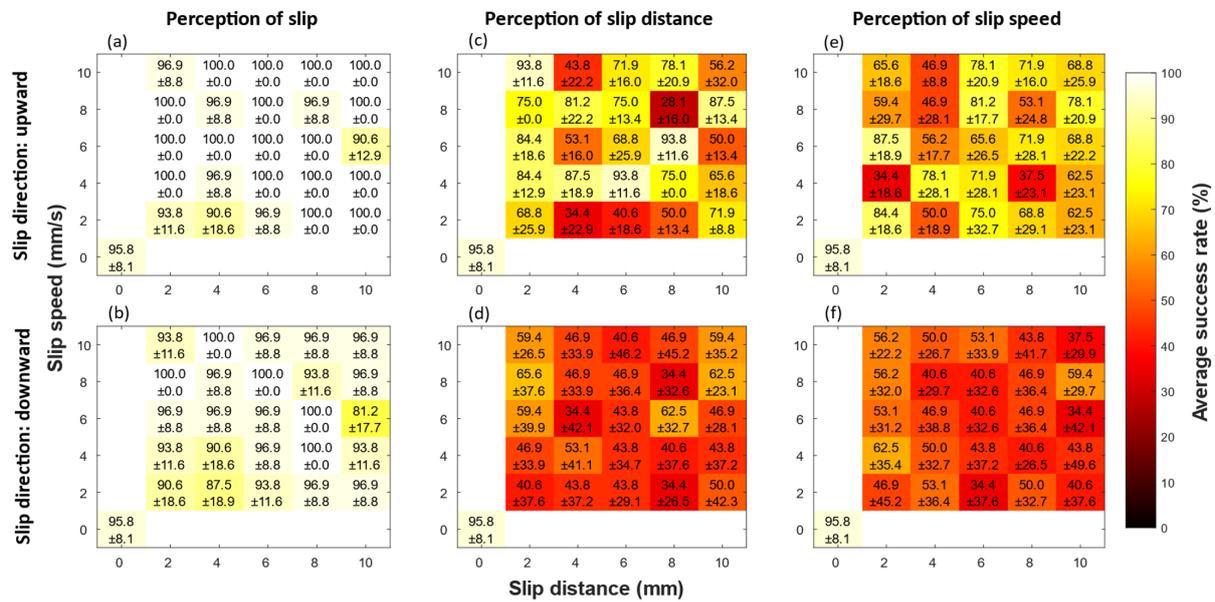

*Figure 2. Heatmaps showing the average success rate for the perception of slip (a and b), perception of slip distance (c and d), and perception of slip speed (e and f). The top row corresponds to the stimulus in the upward direction, and the bottom row corresponds to the downward direction. The average success rate for eight participants is depicted by the color-intensity, and the numerical values mentioned on each cell are the average success rate ± standard deviation.*

**II) Effect of slip direction on the perception of slip distance and slip speed**

Our second hypothesis was validated based on the participant's responses to the second and third psychophysics questions (Q2 and Q3). Firstly, considering the user responses to Q2, Figure 2. (c and d) show the ability of the user to correctly identify differences in slip distance for various combinations of distance and speed stimuli in the upward and downward directions. In the upward slip direction, as observed from Figure 2. (c), participants found perceiving the differences in slip distance more challenging with the lowest slip speed (2mm/s) compared to higher slip speeds (4-10 mm/s). This is evident from the average success rates for the lowest slip speeds which were found to be 34.4±22.9 % (4mm,2mm/s), 40.6±18.6 % (6mm,2mm/s), and 50.0±13.4% (8mm, 2mm/s). Slip distances of 2 mm, 6 mm and 8 mm demonstrate relatively higher discriminability of 93.8±11.6 % (2mm,10mm/s), 93.8±11.6 % (6mm,4mm/s), and 93.8±11.6% (8mm,6mm/s) respectively compared to other slip distances and slip speeds combination in upward direction. The lowest average success rate found for perception of slip distance is 28.1±16% (8mm,8mm/s). On the other hand, considering responses to Q3, Figure 2. (e and f) illustrate the ability to discriminate between differences in slip speed for various combinations of distance and speed stimuli in the upward and downward directions. Results indicate a higher incidence of misidentified trials for smaller distances irrespective of the speeds, with average success rates of 34.4±18.6% (2mm,4mm/s), 46.9±28.1% (4mm,8mm/s), 46.9±8.8% (4mm,10mm/s), and 50.0±18.9% (4mm,2mm/s) in the upward slip direction.

In general, Figure 2. reveals that the average success rates are higher in the upward slip direction as compared to the downward slip direction, whereas the standard deviations are larger in the downward slip direction as compared to the upward slip direction. Furthermore, as observed from Figure 3., the overall average success rate for the perception of slip distance and slip speed in the upward slip direction was found to be 68.5±19.16 % and 65±14.12 %, respectively, while the overall average success rate corresponding to the perception of slip distance (47.88±9.05 %), as well as slip speed (46.88±7.49 %), was drastically reduced in the downward slip direction. Moreover, the three-way interaction between the slip direction, slip distance and slip speed was found to be statistically significant for the perception of difference in slip distances (F (16,112) = 2.287, p = 0.006) and slip speeds (F (16,112) = 2.739, p = 0.001) which suggest that the two-way interactions between slip distance and slip speed are different in the upward and the downward direction.

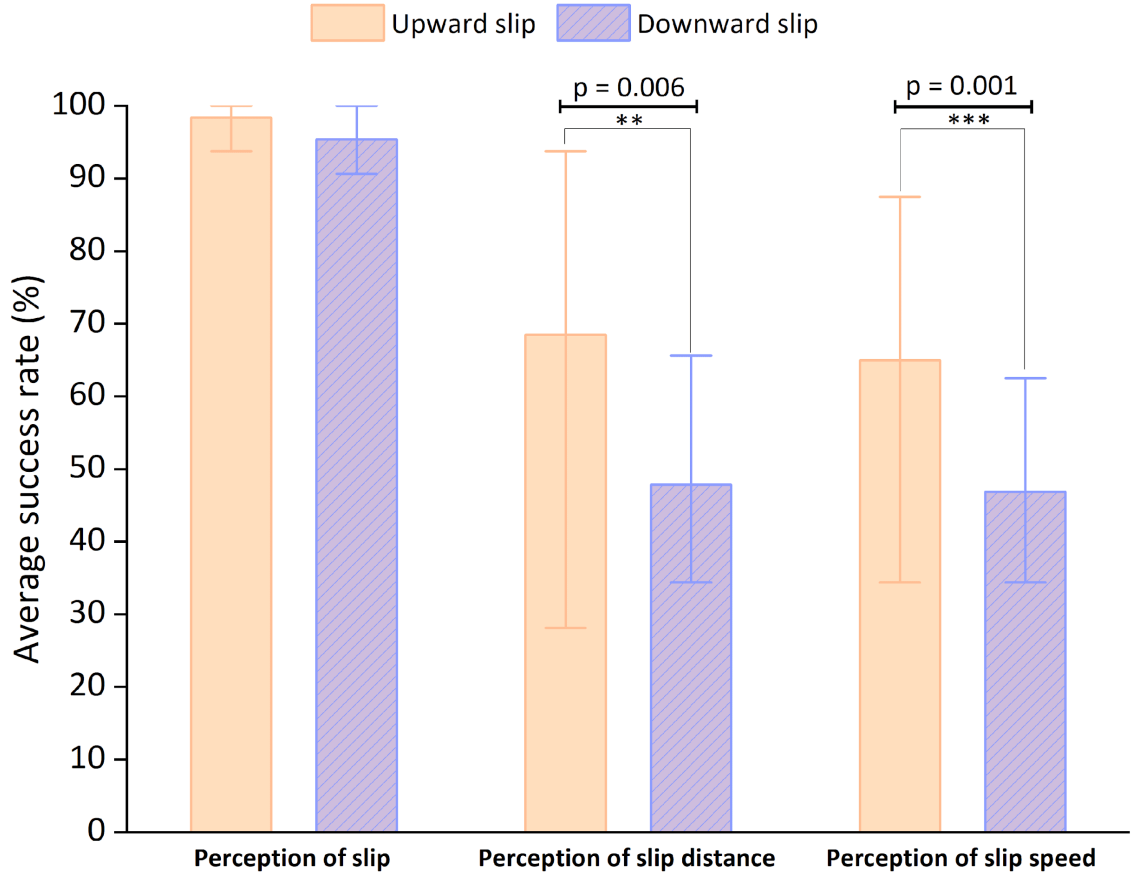

*Figure 3. Comparison of average success rates in the perception of absolute slip, perceptual judgement of slip distance and slip speed. Results from three-way repeated measures ANOVA show statistically significant interaction for the perception of slip distance and slip speed in the upward and downward slip direction.*

### III) Perception of slip distance and slip speed in the upward slip direction

We further investigated if there are any statistically significant simple two-way interactions between slip distance and slip speed for both the slip directions, viz., upward and downward. The simple two-way interaction between slip distance and slip speed was found to be statistically significant for the perception of difference in slip distances ($F(16,112) = 8.679$, $p < 0.001$) and slip speeds ($F(16,112) = 3.764$, $p < 0.001$) of the current and previous slip stimuli in the upward slip direction but not in the downward slip direction ($F(16,112) = 1.194$, $p = 0.284$ for slip distance difference and $F(16,112) = 1.214$, $p = 0.268$ for slip speed difference). This further opens the avenue to investigate the effect of slip speed on the average success rate of perception of slip distance and slip speed for each slip distance in the upward slip direction.

### IV) Interaction between slip distances and slip speeds in the upward slip direction

Table I: Interaction between slip distances and slip speeds in the upward slip direction.

| Slip distance (mm) | Perception of slip distance ($\alpha = 0.05$) | Perception of slip speed ($\alpha = 0.05$) |
|---|---|---|
| 2 | $F(4,28) = 2.956$, $p = 0.037$ | $F(1.868,13.078) = 11.162$, $p = 0.002$ |
| 4 | $F(4,28) = 11.236$, $p < 0.001$ | $F(4,28) = 6.487$, $p < 0.001$ |
| 6 | $F(4,28) = 7.702$, $p < 0.001$ | $F(4,28) = 0.495$, $p = 0.739$ |
| 8 | $F(4,28) = 29.870$, $p < 0.001$ | $F(4,28) = 4.088$, $p = 0.010$ |
| 10 | $F(4,28) = 5.957$, $p = 0.001$ | $F(4,28) = 0.697$, $p = 0.601$ |

We determined the simple simple main effects to see if slip speed led to different average success rates at different slip distances in the upward slip direction. The average success rate for the perception of slip distance was found to have a statistically significant simple simple main effect of slip speed for all slip distances (Refer Table I) statistically significant simple simple main effect of slip speed for all slip distances except 6mm ($F(1,7) = 0.322$, $p = 0.588$) and 10mm ($F(1,7) = 0.876$, $p = 0.380$).

To further investigate which slip speeds differed from each other on their mean average success rate, simple simple comparisons between different slip speeds were performed. All possible pairwise comparisons were run between different slip speeds for all slip distances for the perception of slip distance as well as the perception of slip speed. Statistically significant pairwise comparisons have been reported in supplementary information (SI Table I and SI Table II). These results indicate a significant mean difference in average success rate at all slip distances for the perception of slip distance; however, for the perception of slip speed, slip distances of 2 mm and 8mm exhibited a significant mean difference in average success rate (Refer supplementary information).

# IV. Discussion

In this study, we investigated how humans perceive slip in the vertical direction while using the prehension grip. We designed the experiment to determine the effect of slip direction on the perception of slip, slip distance, and slip speed. In essence, we investigated if the same slip distance and slip speed are perceived differently in the upward and downward slip direction. We also examined the interaction existing between slip directions, slip distances, and slip speeds. We conducted a psychophysical experiment on eight healthy participants and recorded their subjective responses, assessing their ability to perceive slips in upward and downward directions.

**I) Perception of slip is independent of the slip direction.**

We hypothesized based on existing literature [6], that the ability to perceive slip would not be affected by the slip direction. Our experimental results confirmed our hypothesis that slip direction does not modulate the ability to perceive slip. Although in our study, the slip was simulated in the vertical direction and the participant had grasped the sliding plates of the device with a prehension grasp, the direction of the loading is similar to that of Delhaye et al. [6], i.e., tangential to the finger ridges. This might have produced a similar sensation of the slip at the fingertips and hence might have been perceived in a similar manner in both studies.

**II) Slip direction modulates the perception of slip distance and slip speed**

We also hypothesized that the same slip distance and slip speed will be perceived differently in the upward and the downward slip direction. Our results have demonstrated that the perception of slip distance and slip speed are significantly different in the upward and the downward direction. In essence, the downward slip seems to be more unperceivable, irrespective of the magnitudes of the slip stimulus. Specifically, more unidentified downward slip stimuli exist with lower slip speeds (2mm/s and 4mm/s) and smaller slip distances (2mm and 4mm) across subjects. Moreover, there were more misidentified slip stimuli in the downward slip as compared to the upward slip. Figure 2. also shows that the effect of direction is more prominent in the case of differentiating the magnitudes of slip distance and slip speed of the current and the previous stimulus intensities. Irrespective of the slip direction, the participants seem to find more difficulty differentiating the slip speeds than the slip distances of the current and the previous stimuli. The uncertainty of correct responses in differentiating the slip characteristics among the participants is much higher in the case of a downward slip, as the standard deviation across the participants during a downward slip is much higher than in an upward slip. In essence, the higher values of standard deviation across the participants in the downward slip while differentiating the slip characteristics suggest that it was more difficult for the participants to compare the slip characteristics due to variation in the direction. We believe that the difference in perception of slip parameters for the downward direction could be due to three factors:

*(1) Effect of gravity:* According to our results, the perception of the slip speed and slip distance were accurate in the upward direction but not in the downward direction. We believe that the perception in the downward direction might have some implications for the gravitational effect. Toma et al. demonstrated that the internal representations of gravity in normal and altered gravitational contexts are modulated depending on the input from the mechanoreceptors [10]. They investigated the effect of spatial features of congruent and incongruent visual feedback movement with the actual object displacement in the horizontal and vertical direction was investigated thoroughly. They reported that the grip force was found to be affected by incongruent visual feedback in the vertical direction, which suggests that spatial features of visual feedback motion predict the load force changes [10]. In our setup, visual feedback of the movement was removed, which could have affected the ability to discriminate similar stimuli in upward and downward directions.

*(2) Differences in direction-based perception of cutaneous vibrations:* Object interactions and slip produce cutaneous vibrations that emanate from the point of contact and travel along the hand [11]. Pra et al. have found that the vibrations induced in the tangential direction are perceived more easily than the vibrations induced in the normal direction [12]. Hence, the direction of skin stretch-induced due to the movement of the slip device could have been perceived differently depending on the direction of propagation of the cutaneous vibrations.

*(3) Limitations of experimental design:* We believe that certain limitations are associated with the design of the slip induction device and the experiment protocol. Our results have shown that slip perception in itself is independent of the direction. We believe that the vibrations induced in the device due to motor movement provided tactile cues to the participant during slip simulation. This might have led to biased responses for Q1, especially for the smallest slip distances and lowest slip speeds. Also, we speculate that the participants might have considered the duration of the slip stimulus as the deciding factor while responding to Q2 and Q3. Hence, smaller distances with lower speeds might have been confused with larger distances with higher speeds. The participants' response to Q1 was absolute, whereas they had to memorize the previous stimulus to compare it with the current stimulus to respond to Q2 and Q3. This might have increased the uncertainty of the subjective responses. It would be interesting to see if any significance of the previous stimulus exists in the current stimulus's responses.

Another important aspect of the experiment paradigm is that the stimulus with specific slip characteristics was first provided in the upward direction and then in the downward direction. The results obtained from the experiment have raised an obvious question: Would providing the slip stimulus with specific slip characteristics first in the downward and then in the upward direction alter the results?

**III) Future scope**

As part of this study in future, we wish to investigate the neuromuscular as well as neurological manifestation of slip perception and slip detection to further investigate and correlate the psychophysical findings. These studies may lead to effective closed-loop control strategies for assistive devices for individuals with neuromuscular deficits such as stroke and cerebral palsy.

# Conclusions

The present study exhibited evidence that slip direction modulates the interaction between slip distance and slip speed. It was found that the perception of slip is independent of slip direction, but the perception of slip distance and slip speed does depend on slip direction. The novel contribution of the research includes understanding the psychophysics of slip perception while considering various slip characteristics, viz., direction, distance, and speed. This is the first study which simulates the vertical slip and explores the manifestation of slip on prehension grasp, which closely mimics the real-life scenario. Our research further opens the avenue to investigate the effect of slip direction alongside the associated neuromuscular synergies of prehension grasp.

# Statements and Declaration

## Funding

This work was supported in part by the Science & Engineering Research Board, Department of Science and Technology, Government of India through a Core Research Grant (CRG/2021/004967, PI: B. Mukherjee) and by IIT Delhi through a New Faculty Seed Grant (PLN12/03BM, PI: B. Mukherjee).

## Competing interests

The authors declare that they have no known competing financial or non-financial interests that could have appeared to influence the work reported in this paper.

## Author contributions

All authors contributed to the study conception and design. Data collection and preparation were performed by Ayesha Tooba Khan. Data analysis was performed by Ayesha Tooba Khan, Deepak Joshi and Biswarup Mukherjee. The first draft of the manuscript was written by Ayesha Tooba Khan and, all authors commented on previous versions of the manuscript. All authors read and approved the final manuscript.

## Ethics approval

This study was performed in line with the principles of the Declaration of Helsinki. Approval was granted by the Ethics Committee of the Indian Institute of Technology Delhi, New Delhi, India (Date. August 10, 2022. /No. 2021/P052).

## Consent to publish

Informed consent was obtained from all individual participants included in the study.

## Consent to participate

The authors affirm that human research participants provided informed consent for publication of the image in Figure 1.